\documentclass[aps,pra,preprint,showpacs,showkeys,nofootinbib]{revtex4}
\usepackage{amsfonts}
\usepackage{amsmath}
\usepackage{amssymb}
\usepackage{graphicx}%
\usepackage{epsfig}
\usepackage{color}
\providecommand{\U}[1]{\protect\rule{.1in}{.1in}}

\newcommand{\rmd}{\mathrm{d}}
\newcommand{\rmi}{\mathrm{i}}
\newcommand{\rme}{\mathrm{e}}

\begin{document}
\preprint{ }
\title{High-order harmonic generation from inhomogeneous fields}
\author{M. F. Ciappina$^{1}$}
\author{J. Biegert$^{1,2}$}
\author{R. Quidant$^{1,2}$}
\author{M. Lewenstein$^{1,2}$}
\affiliation{$^{1}$ICFO-Institut de Ci\`ences Fot\`oniques, 08860 Castelldefels (Barcelona), Spain}
\affiliation{$^{2}$ICREA-Instituci\'o Catalana de Recerca i Estudis Avan\c{c}ats, Lluis Companys 23, 08010 Barcelona, Spain}

\begin{abstract}
We present theoretical studies of high-order harmonic generation (HHG) produced by non-homogeneous fields as resulting from the illumination of plasmonic nanostructures with a short laser pulse. We show that both the inhomogeneity of the local fields and the confinement of the electron movement play an important role in the HHG process and lead to the generation of even harmonics and a significantly increased cutoff, more pronounced for the longer wavelengths cases studied. In order to understand and characterize the new HHG features we employ two different approaches: the numerical solution of the time dependent Schr\"odinger equation (TDSE) and the semiclassical approach known as Strong Field Approximation (SFA). Both approaches predict comparable results and show the new features, but using the semiclassical arguments behind the SFA and time-frequency analysis tools, we are able to fully understand the reasons of the cutoff extension.     
\end{abstract}
\maketitle

\section{Introduction}

Coherent light sources in the ultraviolet (UV) to extreme ultraviolet (XUV) spectral range are in high demand nowadays for basic research, material science, biology and possibly lithography~\cite{misharmp}. Their caveat is a demanding infrastructure for XUV generation and target delivery as well as its low efficiency and low duty cycle. The recent demonstration based on surface plasmon resonances as light enhancers could provide a possible solution to this problem~\cite{kim}. 

Field enhanced high-order harmonic generation (HHG) using plasmonics, generated starting from engineered nanostructures, requires no extra cavities or laser pumping to amplify the pulse power. By exploiting surface plasmon resonances, local electric fields can be enhanced by more than 20dB~\cite{muhl,schuck}. This amplification is strong enough to exceed the threshold laser intensity for HHG generation in noble gases and the pulse repetition rate remains unaltered without any extra pumping or cavity attachment. Furthermore, the high harmonics radiation generated from each nanostructure acts as a point-like source, enabling collimation or focusing of this coherent radiation by means of (constructive) interference. This opens a wide range of possibilities to spatially arrange nanostructures to enhance or shape spectral and spatial properties in numerous ways.

The basic principle of high-order harmonic generation (HHG) based on plasmonics can be summarized as follows (the full explanation can be found in~\cite{kim}): a femtosecond low intensity pulse is coupled to the plasmon mode inducing a collective oscillation of free charges within the metal. The free charges redistribute the electric
field around each of the nanostructure, thereby forming a spot of highly enhanced electric field. The enhanced field exceeds the
threshold of HHG, thus by injection of noble gases onto the spot of the enhanced field, high
harmonics are generated.

In the seminal experiment of Kim et al.~\cite{kim}, the output beam emitted from a modest power femtosecond
oscillator (100-kW peak power, 1.3 nJ pulse energy, 10 fs pulse duration and 800 nm of
wavelength) was directly focused onto a 10 $\mu$m$\times$10 $\mu$m bow-tie nanoantenna array with a pulse intensity
of 10$^{11}$ W$\cdot$cm$^{-2}$, which is about two orders of magnitude less than the threshold intensity to
generate HHG in noble gases. Their experimental result showed that the field intensity
enhancement factor exceeded 20 dB, which is sufficient to produce XUV wavelengths
from the 7th (114 nm) to the 21st (38 nm) harmonics with the injection of xenon gas. Additionally, each bow-tie
acts as a point-like source, thus 3D arrangement of bow-ties should enable us to perform control of generated harmonics in various ways by exploiting interference.

Numerical and semiclassical approaches to study laser-matter processes in atoms and molecules, in particular High-order Harmonic Generation (HHG), are largely based on the assumption that the laser electric field is homogeneous in the region where the electron dynamics takes place~\cite{keitel,krausz}. Due to the strong confinement of plasmonic hot spots, the laser electric field is clearly not homogeneous anymore, in the region where the electron dynamics takes place, and consequently important changes in the laser matter processes would occur. From a theoretical viewpoint, the HHG process can be tackled using different approaches (for a summary see e.g.~\cite{book1,book2} and references therein). In this article we concentrate our effort in extending two of the most and widely used approaches: the numerical solution of Time Dependent Schr\"odinger Equation (TDSE) in reduced dimensions and the Strong Field Approximation~\cite{sfa}. So far theoretical work in strong field physics has been focused on the processes driven
by homogeneous coherent electromagnetic radiation on atoms and molecules. Only recently
studies about how HHG spectra are modified due to non-homogeneous fields, as those
present in the vicinity of a nanostructure irradiated by a short laser pulse, have been
published~\cite{husakou,yavuz}.

The paper is organized as follows. In the next two sections (Sec. II.A and Sec. II.B) we will present two theoretical approaches to model high-order harmonic generation (HHG) spectra produced by non-homogeneous fields, namely the numerical solution of the Time Dependent Schr\"odinger Equation in reduced dimensions (1D-TDSE) and the Strong Field Approximation (SFA), respectively. We have developed both approaches in such a way to allow the treatment of very general non-homogeneous fields showing the flexibility of both the 1D-TDSE and SFA models. Particular studies of HHG spectra for the most simplest case is presented in Sec.  III. We will discuss how the non-homogeneous field produce noticeable modifications in the HHG spectra, namely a change in the harmonic periodicity (odd and even harmonics now appear) and a distinct extension in the position of the cutoff (more pronounced for the longer wavelengths cases studied).  Both the 1D-TDSE and SFA approaches provide comparable predictions, but analyzing the HHG process using semiclassical arguments we can present strong evidence about the motives of the cutoff extension. The paper ends with a short summary and an outlook.

\section{Theory}

\subsection{Numerical solution of the Time Dependent Schr\"odinger Equation in reduced dimensions  (1D-TDSE) }
\indent Considering the dynamics of an atomic electron in a strong laser field is mainly along the direction of the field (in a linearly polarized laser pulse), it is reasonable to model the high-order harmonic generation (HHG) in a 1D spatial dimension by solving the following Schr\"odinger equation (1D-TDSE):
\begin{eqnarray}
\label{tdse}
\rmi \frac{\partial \Psi(x,t)}{\partial t}&=&\mathcal{H}(t)\Psi(x,t) \\
&=&\left[-\frac{1}{2}\frac{\partial^{2}}{\partial x^{2}}+V_{atom}(x)+V_{laser}(x,t)\right]\Psi(x,t) \nonumber
\end{eqnarray}
To model an atom in 1D, it is common to use the quasi-Coulomb potential
\begin{eqnarray}
\label{atom}
V_{atom}(x)&=&-\frac{1}{\sqrt{x^2+1}}
\end{eqnarray}
which was first introduced in~\cite{eberly} and has been widely used in the 1D analysis of an atom. The
potential due to the laser electric field linearly polarized along the x-axis will be modified in order to treat nonhomogeneous fields. Consequently we write
\begin{eqnarray}
\label{vlaser}
V_{laser}(x,t)&=&E(x,t)\,x
\end{eqnarray}
with 
\begin{equation}
\label{electric}
E(x,t)=E_0\,f(t)\, (1+\varepsilon g(x))\,\sin\omega t
\end{equation}
the laser electric field. In (\ref{electric}) $E_0$ is the peak amplitude and $\omega$ the frequency of the coherent electromagnetic radiation. Furthermore, $f(t)$ defines the pulse envelope and $\varepsilon$ a small parameter that characterizes the inhomogeneity region. $g(x)$ represents the functional form of the nonhomogeneous field. In this work we concentrate our analysis in the simplest form for $g(x)$, i.e. $g(x)=x$, but we should to emphasize that the numerical algorithm allows, in principle, to use any functional form for $g(x)$. We note they for the particular case $g(x)=x$ $\varepsilon$ has dimensions of inverse length. To model the laser pulses, we shall use a trapezoidal envelope $f(t)$ given by
\begin{equation}
\label{shape}
  f(t) = \left\{
  \begin{array}{l l}
    \frac{t}{t_{1}} & \quad \text{for $0 \leq t < t_1$}\\
    1 & \quad \text{for $t_1 \leq t \leq t_2$}\\
    -\frac{(t-t_3)}{(t_{3}-t_{2})} & \quad \text{for $t_2 < t \leq t_3$}\\
    0 & \quad \text{elsewhere}\\
  \end{array} \right.
\end{equation}
where $t_1=2\pi n_on/\omega$, $t_2=t_1+2\pi n_{p}/\omega$, and $t_3=t_2+2\pi n_{off}/\omega$. $n_{on}$, $n_p$ and $n_{off}$ are the number of cycles of turn on, plateau and turn off, respectively.

The initial state in the 1D-TDSE is the ground state (GS) of the system before we turn on the laser ($t=-\infty$) and it can be found solving an eigenvalue problem once the spatial coordinate $x$ has been discretized. The corresponding eigenvalue for the potential (\ref{atom}) is found to be
$\mathcal{E}_{GS} = -18.2$ eV ($-0.67$ a.u.). For comparison, we note that the ground state energy of Ne is $-21.6$ eV ($-0.79$ a.u.), and $-15.8$ eV ($-0.58$ a.u.) for Ar.

Equation (\ref{tdse}) can be solved numerically by using the Crank-Nicolson scheme~\cite{keitel}. In order to avoid spurious reflections from the boundaries, at each time step, the total wave function is multiplied by a mask function of the form $cos^{1/8}$, which varies from 1 to 0 starting from the 2/3 of the grid~\cite{mask}.

Once having found the state $\Psi(x,t)$ of the system from the 1D-TDSE (\ref{tdse}), we can calculate the harmonic spectrum as follows~\cite{schafer}. The harmonic yield of an atom is proportional to the Fourier transform of the acceleration $a(t)$ of its active electron. That it,
\begin{equation}
D(\omega)=\left| \frac{1}{\tau}\frac{1}{\omega^2}\int_{-\infty}^{\infty}\rmd t\rme^{-\rmi \omega t}a(t)\right|^2
\end{equation}
where $a(t)$ can be obtained by using he following commutator relation
\begin{equation}
\label{accel1D}
a(t)=\frac{\rmd^{2}\langle x \rangle}{\rmd t^2}=-\langle \Psi(t) | \left[ \mathcal{H}(t),\left[ \mathcal{H}(t),x\right]\right] | \Psi(t) \rangle,
\end{equation}
where $\mathcal{H}(t)$ is the Hamiltonian defined in the Eq. (\ref{tdse}). The function $D(\omega)$
is called the dipole spectrum, since $D(\omega)$ gives the spectral profile measured in HHG experiments.

\subsection{The Strong Field Approximation (SFA) for inhomogeneous fields}
\indent  Another model to evaluate high-harmonic spectra for atoms in intense laser pulses is the Lewenstein model~\cite{sfa}.  The main ingredient of this approach is the evaluation of the time dependent dipole moment $\mathbf{d}(t)$. Within the Single Active Electron (SAE) approximation and considering the harmonic radiation is directed mainly in the $x$-axis $\mathbf{d}(t)$ in the length gauge for an atom can be written~\cite{sfa}
\begin{eqnarray}
\label{dipolesfa}
d_x(t)&=&-\rmi\int_{t_0}^{t}\rmd t'\int \rmd \mathbf{k}\, d_{ion,x}(\mathbf{k}+\mathbf{A}(t'),t')\nonumber\\
&\times& d_{rec,x}^{*}(\mathbf{k}+\mathbf{A}(t))\exp\left[-\rmi S_0(\mathbf{k},t,t')\right]+c.c
\end{eqnarray}
In (\ref{dipolesfa})
\begin{equation}
\label{s0}
S_0(\mathbf{k},t,t')=\int_{t'}^{t}\rmd t'' \left\{\frac{\left[\mathbf{k}+\mathbf{A}(t'')\right]^2}{2}+I_p\right\}
\end{equation} 
is the semiclassical action, $I_p$ is the ionization potential of the atom, and $\mathbf{A}(t)=-\int_{-\infty}^{t}\mathbf{E}(t')\rmd t$ is the vector potential of the linearly polarized laser electric field $\mathbf{E}(t)$. The ionization and recombination matrices are given by
\begin{eqnarray}
d_{ion,x}(\mathbf{k},t)&=&\langle \Psi_{\mathbf{k}}|E(t)\, x |\phi_0\rangle 
\end{eqnarray}
and
\begin{eqnarray}
d_{rec,x}(\mathbf{k})&=&\langle \Psi_{\mathbf{k}}|-x |\phi_0\rangle 
\end{eqnarray}
respectively. $\Psi_{\mathbf{k}}$ is a normalized plane wave of momentum $\mathbf{k}$
\begin{eqnarray}
\label{pw}
\Psi_{\mathbf{k}}(\mathbf{r})&=&(2\pi)^{-3/2}\rme^{\rmi\mathbf{k}\cdot\mathbf{r}} 
\end{eqnarray}
and $\phi_0$ is the undressed initial-state of the atom. In our studies we use hydrogenic $1s$ states of the form
\begin{eqnarray}
\label{ground}
\phi_0(\mathbf{r})&=&\sqrt{\frac{\lambda^{3}}{\pi}}\rme^{-\lambda r} 
\end{eqnarray}
and we chose the effective charge $\lambda$ in order to match the energy of the ground state $\mathcal{E}_{GS}$ of the 1D model atom (see Sec. II.A), i.e. $\lambda=\sqrt{2|\mathcal{E}_{GS}|}$. Using (\ref{pw}) and (\ref{ground}) the explicit expressions for $d_{ion,x}(\mathbf{k})$ and $d_{rec,x}(\mathbf{k})$ are
\begin{eqnarray}
d_{ion,x}(\mathbf{k},t)&=&\rmi\frac{2^{7/2}\lambda^{5/2}}{\pi}\frac{k_z}{(\mathbf{k}^2+\lambda^2)^{3}}E(t)
\end{eqnarray}
and
\begin{eqnarray}
d_{rec,x}(\mathbf{k})&=&-\rmi\frac{2^{7/2}\lambda^{5/2}}{\pi}\frac{k_z}{(\mathbf{k}^2+\lambda^2)^{3}}
\end{eqnarray}
respectively. The spectrum of the emitted light polarized along the x-axis is obtained by modulus squaring the Fourier 
transform of the dipole acceleration
\begin{equation}
a_{x}(\Omega)=\int_{0}^{T_p}\rmd t \ddot{d}_x(t)\exp(\rmi\Omega t)
\end{equation}
where the integration is carried out over the duration of the
laser pulse, $T_p$, by applying a fast Fourier transform algorithm.
The numerical calculation of Eq. (\ref{dipolesfa}) involves
a multidimensional integration over momentum and time. As
usual~\cite{sfa}, we have performed the three-dimensional integration
over $\mathbf{k}$ using the saddle point or stationary phase
method, meanwhile all time integrations are performed numerically.

Eq. (\ref{dipolesfa}) has a direct interpretation in terms of the three step or simple man's model~\cite{sfa}. The first step is the strong-field ionization
of the atom or molecule as a consequence of the nonperturbative interaction with the coherent electromagnetic radiation.
The classical propagation of the electron in the field defines the second step of the model. Finally, the third step in
the sequence occurs when the electron is steered back in the linearly polarized field to its origin, recombining under the
emission of a high-energy photon. One of the main features of the HHG process is the coherence of the emitted radiation,
which, e.g., opens the possibility of generating attosecond pulses~\cite{corkumnat}.

The Lewenstein model implicitly considers homogeneous laser electric fields, i.e.the electric field does not change in the region where the electron dynamics takes place. In order to consider non-homogeneous fields, the SFA needs to be modified accordingly. Our goal is to find the modifications in the electron momentum and the classical action produced by the non-homogeneous  field. The first step is to find the electron trajectories starting from classical arguments employing the Newton equation for an electron moving in a non-homogeneous electric field $E(x,t)$, linearly polarized in the x-axis. It can be then written as
\begin{eqnarray}
\label{newton}
\nonumber
\ddot{x}(t)&=&-\nabla_x V_{laser}(x,t)\\
&=&-E(x,t)-\left[\nabla_x E(x,t)\right] x \nonumber\\
&=&-E(t)(1+2\varepsilon x(t))
\end{eqnarray} 
where $V_{laser}(x,t)$ is given by (\ref{vlaser}) and we have collected the time dependent part of the electric field in $E(t)$, i.e $E(t)=E_0 f(t) \sin\omega t$.
We use the Picard iteration~\cite{diffeq} extended to second order ordinary differential equations to solve Eq.(\ref{newton}) and we restrict ourselves only to the first order term. The ($n+1$)th order solution can be written in terms of the $n$th one as follows
\begin{eqnarray}
X_{n+1}(t)&=&x_0+v_0(t-t_0) \\
&+&\int_{t_0}^{t}\left[\int_{t_0}^{t'}f(t'',X_{n}(t''),\dot{X}_n(t''))\rmd t''\right]\rmd t' \nonumber
\end{eqnarray} 
 where $x_0=x(t_0)$, $v_0=\dot{x}_0=\dot{x}(t_0)$ and in our case $f(t,X_{n}(t),\dot{X}_n(t))=-E(t)(1+2\varepsilon X_{n}(t))$, $E(t)$ being the time dependent part of $E(x,t)$, i.e. $E(t)=E_0 f(t) \sin \omega t$. Considering the initial conditions, i.e. $x_0=0$ and $v_0=0$ (the electron starts its movement at the origin with zero velocity), we finally obtain
\begin{eqnarray}
\label{x1}
X_1(t)&=&\alpha(t)-\alpha(t_0)-A(t_0)(t-t_0)
\end{eqnarray} 
where we have defined $\alpha(t)=\int_{0}^{t}\rmd t' A(t')$. The next step is to calculate the classical action and the saddle point electron momentum starting from the electron trajectories. After elementary algebra we can write for the modified action $S(\mathbf{k},t,t')$
\begin{eqnarray}
\label{s}
S(\mathbf{k},t,t')&=&S_0(\mathbf{k},t,t') \nonumber\\
&+&2\varepsilon\left[\mathbf{k}\cdot\int_{t'}^{t}\rmd t''\mathbf{A}(t'')X_{1}(t'')\right. \\
&+&\left.\int_{t'}^{t}\rmd t''\mathbf{A}^{2}(t'')X_{1}(t'')\right] \nonumber
\end{eqnarray} 
where $S_0(\mathbf{k},t,t')$ is defined in (\ref{s0}) and $X_{1}(t)$ is the electron trajectory of Eq. (\ref{x1}). The saddle point momentum is found from the stationary condition
\begin{equation}
\nabla_{\mathbf{k}}S(\mathbf{k},t,t')=0
\end{equation}
Consequently
\begin{equation}
\nabla_{\mathbf{k}}S(\mathbf{k},t,t')=\nabla_{\mathbf{k}}S_0(\mathbf{k},t,t')+2\varepsilon \int_{t'}^{t}\rmd t''\mathbf{A}(t'')X_{1}(t'')
\end{equation}
with
\begin{equation}
\nabla_{\mathbf{k}}S_0(\mathbf{k},t,t')=\mathbf{k}(t-t')+\alpha(t)-\alpha(t')
\end{equation}
Finally we obtain for the saddle point or stationary momentum $\mathbf{k}_{st}(t,t')$
\begin{eqnarray}
\label{kst}
\mathbf{k}_{st}(t,t')&=&-\frac{\alpha(t)-\alpha(t')}{(t-t')} \nonumber\\
&-&\frac{2\varepsilon}{(t-t')}\int_{t'}^{t}\rmd t''\mathbf{A}(t'')X_{1}(t'')
\end{eqnarray}
Replacing (\ref{kst}) in (\ref{s}) results
\begin{eqnarray}
\label{sfinal}
S(\mathbf{k}_s,t,t')&=&S_0(\mathbf{k}_s,t,t')\nonumber\\
&+&2\varepsilon\left[\mathbf{k}_{st}\cdot\int_{t'}^{t}\rmd t''\mathbf{A}(t'')X_{1}(t'')\right.\\
&+&\left.\int_{t'}^{t}\rmd t''\mathbf{A}^{2}(t'')X_{1}(t'')\right]\nonumber
\end{eqnarray} 
The time dependent dipole moment, Eq. (\ref{dipolesfa}), can be written, after the saddle point method for the momentum $\mathbf{k}$ is applied~\cite{sfa}, as
\begin{eqnarray}
\label{dipolesfa2}
d_x(t)&=&-\rmi\int_{t_0}^{t}\rmd t'\left(\frac{\pi}{\eta+\rmi(t-t')/2}\right)^{3/2}\nonumber\\
&\times& d_{ion,x}(\mathbf{k}_{st}(t,t')+\mathbf{A}(t'),t')\\
&\times& d_{rec,x}^{*}(\mathbf{k}_{st}(t,t')+\mathbf{A}(t))\exp\left[-\rmi S(\mathbf{k}_{st},t,t')\right]+c.c \nonumber
\end{eqnarray} 
where $\eta$ is a small parameter. We recover the \textit{homogeneous} dipole moment putting $\varepsilon=0$ in (\ref{dipolesfa2})~\cite{sfa}.

\section{Results}
\indent Results using the 1D-TDSE model developed in Sec. II.A are presented in Figs. 1-7.  In Figs. 1-6 the laser intensity is $I=2\times10^{14}$ W$\cdot$cm$^{-2}$ and the laser wavelength is $\lambda=800$ nm. We have used a trapezoidal shaped pulse with two optical cycles turn on, i.e. $n_{on}=2$, and turn off, i.e. $n_{off}=2$, and a plateau with six optical cycles, i.e. $n_{p}=6$ (10 optical cycles in total, i.e. approximately 27 fs). The model atom has $\mathcal{E}_{GS}=-0.67$ a.u. and it is driven by a laser electric field of the form $E(x,t)=E_0 f(t)(1+\varepsilon x(t))$, where $E_0$ is the peak amplitude and $f(t)$ is the pulse envelope (see panel (e) of Fig. 2 for a plot of the laser electric field).

%\begin{figure}[htb]
%\centering
%\includegraphics[width=0.4\textwidth]{figure1.eps}
%\caption{High-order harmonic generation (HHG) spectra for a model atom with $\mathcal{E}_{GS}=-0.67$ a.u. generated using the 1D-TDSE model and with a spatial grid of $x_{lim}=\pm7.5 \alpha_0$ (see text for details). The laser parameters are $I=2\times10^{14}$ W$\cdot$cm$^{-2}$ and $\lambda=800$ nm. We have used a trapezoidal shaped pulse with two optical cycles turn on, i.e. $n_{on}=2$, and turn off, i.e. $n_{off}=2$, and a plateau with six optical cycles, i.e. $n_{p}=6$ (10 optical cycles in total, i.e. approximately 27 fs). The arrow indicates the cutoff predicted by the semiclassical model~\cite{sfa}. Panel (a) homogeneous case, (b) $\varepsilon=0.01$ (100 a.u), (c) $\varepsilon=0.02$ (50 a.u) and (d) $\varepsilon=0.05$ (20 a.u).}
%\label{fig:figure1}
%\end{figure}

One additional parameter we take into account in the 1D-TDSE simulations is the spatial region where the electron moves. This parameter will naturally appear in real situations using bow-tie-shaped nanostructures, as those employed in the experiments of Kim et.al~\cite{kim}, and when linearly polarized pulses are utilized, that restricts the electron dynamics mostly to one dimension. The bow-tie shaped nanostructures are characterized by a spatial gap that can be, in principle, changed between certain ranges when the nanostructure is engineered. We use in our 1D-TDSE model spatial grids in terms of the quiver radius, $\alpha_0=E_0/\omega^{2}$, $\omega$ being the driven laser frequency, ($\omega=0.057$ a.u. for a laser wavelength $\lambda=800$ nm) that in the case under study is $\alpha_0\approx 23.2$ a.u. (1.23 nm). Three different spatial grid sizes will be employed, namely $x_{lim}=\pm7.5\alpha_0$, $x_{lim}=\pm4.5\alpha_0$ and $x_{lim}=\pm1.5\alpha_0$, corresponding to gaps of 18.7 nm, 11 nm and 3.7 nm, respectively . 

%\begin{figure}[htb]
%\centering
%\includegraphics[width=.85\textwidth]{figure2.eps}
%\caption{Panels (a)-(d): Gabor analysis for the HHG spectra of Figure 1. The zoomed regions in all panels show a time interval during the laser pulse for which the complete electron trajectory, from birth time to recollision time, falls within the pulse plateau (see Ref.~\cite{manfred} for details); panel (e) shape of the laser electric field. In panels (a)-(d) the color scale is logarithmic.}
%\label{fig:figure2}
%\end{figure}

In Fig. 1 we plot the HHG spectra for different values of $\varepsilon$ and for a grid size of $x_{lim}=\pm7.5\alpha_0$. Panel (a) is the homogeneous case and we have chosen values of 0.01 (panel b), 0.02 (panel c) and 0.05 (panel d) for the inhomogeneity parameter $\varepsilon$ that corresponds to inhomogeneity regions of 100 a.u. (5.3 nm), 50 a.u. (2.7 nm) and 20 a.u. (1 nm), respectively. Two distinct characteristics can be observed in panels (b)-(d), and we can summarize them as follows
(i) Odd and even harmonics are now present. This new feature appears because we have broken the symmetry of the total potential, $V_{atom}+V_{laser}$, introducing a new asymmetric term, i.e. $\varepsilon x^2$;
(ii) The absence of a clear HHG cutoff. This effect is related with the electron trajectories that contribute to the harmonic spectra and will be object of study using the Gabor analysis and the semiclassical simulations (see below for details).

Our next step is to use time-analysis tools in order to characterize the HHG spectra calculated using the 1D-TDSE model. To this end we employ the Gabor transformation, developed in the 1940s by D. Gabor~\cite{gabor}, that has shown to be a very powerful tool in order to estimate the emission times of HHG in atoms and molecules and to discriminate the different electron trajectories~\cite{manfred}. Starting from the dipole acceleration $a(t)$ of Eq. (\ref{accel1D}) the Gabor transform is defined as
\begin{eqnarray}
a_{G}(\Omega,t)&=&\int dt' a(t') \frac{\exp\left[-(t-t')^{2}/2\sigma^{2} \right]}{\sigma \sqrt{2 \pi}}\exp(i \Omega t')
\end{eqnarray}
where the integration is usually taken over the pulse duration. In our studies we use $\sigma=1/3 \omega$, with $\omega$ being the central laser frequency. With this value of $\sigma$ we achieve an adequately balance between the time and frequency resolutions  (see Ref.~\cite{manfred} for details). Results of the Gabor analysis of the HHG spectra of Fig. 1 are presented in Fig. 2. From panels (a)-(d) can be observed the clear differences between the homogeneous (panel (a)) and non-homogeneous cases (panel (b)-(d)). In the zoomed regions we show a time interval during the laser pulse for which a complete electron trajectory, from birth time to recollision time, falls within the pulse plateau. From these plots it is possible to clearly observe the short (for emission times smaller than $\sim$320 a.u.) and long (the emission times are larger than $\sim$320 a.u. for this case) trajectories (see Ref.~\cite{manfred} for more details). The highest harmonic order (around the 40th) is generated at an emission time of $\sim$320 a.u. for the homogeneous case. On the other hand, only an \textit{extended} short trajectory (with emission times smaller than $\sim$320 a.u.) is present for all the nonhomogeneous cases and no a clear harmonic cutoff is visible.

%\begin{figure}[htb]
%\centering
%\includegraphics[width=.5\textwidth]{figure3.eps}
%\caption{Idem Fig. 1 but with a spatial grid of $x_{lim}=\pm4.5 \alpha_0$.}
%\label{fig:figure3}
%\end{figure}

Alternatively, in Figs. 3 and 4 we plot the HHG spectra for different values of $\varepsilon$ and the Gabor analysis, respectively. Here a grid size of $x_{lim}=\pm4.5\alpha_0$ is used. Panel (a) is the homogeneous case and we have chosen values of 0.01 (panel b), 0.02 (panel c) and 0.05 (panel d) for the inhomogeneity parameter $\varepsilon$ that corresponds to inhomogeneity regions of 100 a.u. (5.3 nm), 50 a.u. (2.7 nm) and 20 a.u. (1 nm), respectively. We note that the difference between these last two figures (Figs. 3 and 4) and Figs. 1 and 2 is hardly visible for all the cases, showing that the decrease of the grid size has no effect in the HHG spectra, both for the homogeneous and nonhomogeneous cases. 
%
%\begin{figure}[htb]
%\centering
%\includegraphics[width=\textwidth]{figure4.eps}
%\caption{Gabor analysis for the HHG spectra of Figure 3. The zoomed regions in all panels show a time interval during the laser pulse for which the complete electron trajectory, from birth time o recollision time, falls within the pulse plateau (see text and Ref.~\cite{manfred} for details).}
%\label{fig:figure4}
%\end{figure}
%%
%\begin{figure}[htb]
%\centering
%\includegraphics[width=.5\textwidth]{figure5.eps}
%\caption{Idem Fig. 1 but with a spatial grid of $x_{lim}=\pm1.5 \alpha_0$.}
%\label{fig:figure5}
%\end{figure}

Finally, in Figs. 5 and 6 we plot the HHG spectra for different values of $\varepsilon$ and the Gabor analysis, respectively, and now with a grid size of $x_{lim}=\pm1.5\alpha_0$. Panel (a) is the homogeneous case and we have chosen values of 0.01 (panel b), 0.02 (panel c) and 0.05 (panel d) for the inhomogeneity parameter $\varepsilon$ that corresponds to inhomogeneity regions of 100 a.u. (5.3 nm), 50 a.u. (2.7 nm) and 20 a.u. (1 nm), respectively. Here it is possible to observe the substantial difference between panels (b)-(d) of these last two figures and those from Figs. 1-4, showing that the electron confinement, joint with the inhomogeneities of the laser electric field, are the responsible of the appearance of a clear extended harmonic cutoff. We also note the HHG spectra for the homogeneous case (panel (a) of Figs. 1, 3 and 5) are identical for all the cases, confirming that not only the electron confinement is the reason of the new features present in the HHG spectra. In some sense, when we restrict the electron movement, only the short trajectories contribute to the formation of the harmonic spectrum and this feature is clearly visible in the zoomed regions of the Gabor analysis in Fig. 6 where only short trajectories are distinguishable, i.e. those with emission times smaller than $\sim$320 a.u. 

%\begin{figure}[htb]
%\centering
%\includegraphics[width=\textwidth]{figure6.eps}
%\caption{Gabor analysis for the HHG spectra of Figure 5. The zoomed regions in all panels show a time interval during the laser pulse for which the complete electron trajectory, from birth time o recollision time, falls within the pulse plateau (see text and Ref.~\cite{manfred} for details).}
%\label{fig:figure6}
%\end{figure}

In order to explore how our 1D-TDSE approach behaves we have calculated HHG spectra for longer wavelengths. We present our results in Fig. 7 for $\lambda=3.2$ $\mu$m. This value for $\lambda$ represents an example that could be perfectly reachable experimentally~\cite{jens1,jens2} and allow us to reach values for the electron excursion region (i.e. the quiver radius $\alpha_0$) closer to those considered in real experiments~\cite{kim}. The laser intensity is $I=1\times10^{13}$  W$\cdot$cm$^{-2}$ and we have employed a gaussian shaped pulse with 6 cycles FWHM, $f(t)=\exp\left[-\frac{2\ln 2}{\tau^{2}}t^{2}\right]$,where $\tau$ is the FWHM (full-width at half-maximum) duration of the driving laser intensity $I(t)$ ($\propto|E(t)|^2$). We use a spatial grid of $x_{lim}=\pm 1.5 \alpha_0$, where $\alpha_0=83$ a.u. (4.4 nm), i.e. the electron dynamics is confined in a region of around 13.2 nm. The panels correspond to the homogeneous case (a), $\varepsilon=0.01$ (b), $\varepsilon=0.02$ (c) and $\varepsilon=0.05$ (d), respectively. We have included zoomed panels for the (a) and (d) cases. These two plots allow us to observe clearly the presence of odd and even harmonics for this particular nonhomogeneous case and only the appearance of odd harmonics for the homogeneous one. From Fig. 7 we conclude that for longer wavelengths the cutoff extension due to the nonhomogeneities of the field, combined with the electron confinement, is far more pronounced, e.g. for the case of $\lambda=3.2$ $\mu$m and $\varepsilon=0.05$ the cutoff is almost 3 times larger than the homogeneous case. This extension could open the avenue for the production of attosecond pulses in the keV regime (for a theoretical study at $\lambda=800$ nm see~\cite{yavuz}). Furthermore the region of confinement using these laser parameters is closer to values that could be found in real nanostructures. 

%\begin{figure}[htb]
%\centering
%\includegraphics[width=0.6\textwidth]{figure7.eps}
%\caption{ HHG spectra for a model atom with $\mathcal{E}_{GS}=-0.67$ a.u. generated using the 1D-TDSE model  and with a spatial grid of $x_{lim}=\pm1.5 \alpha_0$ (see text for details). The laser parameters are $I=1\times10^{13}$ W$\cdot$cm$^{-2}$ and $\lambda=3.2$ $\mu$m (3200 nm). We have used a gaussian shaped pulse with 6 cycles FWHM. The arrow indicates the cutoff predicted by the semiclassical model~\cite{sfa}. The insets in panels (a) and (d) show particular zoomed regions of the harmonic spectra near the cutoff region (see the text for details).}
%\label{fig:figure7}
%\end{figure}

In the following we use the modified SFA model presented in the Sec. II.B to generate HHG spectra produced by nonhomogeneous fields. We employ a laser pulse with an electric field of the form $E(x,t)=E_0 f(t)(1+\varepsilon x(t))$, where $E_0$ is the peak amplitude, $f(t)$ is the pulse envelope and we use only the first order for $x(t)$, i.e. $x(t)\approx X_1(t)$ with $X_1(t)$ of (\ref{x1}). In Fig. 8 we show the predictions of this model and Fig. 9 represents the Gabor analysis of the HHG spectra of the former. In order to enhance the new HHG features and see more clearly the differences between the studied cases now the laser field has an intensity $I=6\times10^{14}$  W$\cdot$cm$^{-2}$ and a wavelength of $\lambda=800$ nm. We have used a trapezoidal shaped pulse with three optical cycles turn-on ($n_{on}=3$) and turn off ($n_{off}=4$) and a plateau of four constant-amplitude optical cycles ($n_p=4$) (see panel (e) of Fig. 9) and the model atom has $\mathcal{E}_{GS}=-0.67$ a.u. With these parameters the predicted cutoff is $n_c=85$ as it is shown by the arrow plotted in panel (a). Panels (b), (c) and (d) correspond to values of $2\varepsilon=0.01$,  $2\varepsilon=0.02$ and $2\varepsilon=0.05$, respectively. As in the case of the 1D-TDSE model, the SFA approach predicts the appearance of odd and even harmonics and a cutoff extension. In order to show more clearly these features we have zoomed out a small region of the HHG near the cutoff for panels (a) and (b). In this graph it is more transparent to see how the inhomogeneous fields modify the harmonic spectra. An additional feature appears and it could be extracted from panels (a) (homogeneous case) and (d) ($2\varepsilon=0.05$): there exists a clear enhancement in the harmonic yield in the plateau region (see e.g. a region around the 25th-40th harmonics). The Gabor analysis of the HHG spectra calculated using the SFA, Fig. 9, shows similarities and differences between this model and the 1D-TDSE approach (the SFA approach does not account for the restrictions on the electron motion). For example, in panel (c) only short trajectories appear to contribute to the HHG spectra (similar to the $x_{lim}=\pm1.5 \alpha_0$ case), meanwhile panel (d) has analogous features to panels (d) of Figs. 2 and 4 (i.e. the $x_{lim}=\pm7.5 \alpha_0$  and $x_{lim}=\pm4.5 \alpha_0$ cases respectively). This behavior could be related to the limited validity of the perturbation-like character of the SFA approach developed in Sec. II.B. An extension of the SFA approach including the quantum orbits analysis will be published elsewhere.

%
%\begin{figure}[htb]
%\centering
%\includegraphics[width=0.6\textwidth]{figure8.eps}
%\caption{HHG spectra for a model atom with $\mathcal{E}_{GS}=-0.67$ generated using the extended SFA approach. The laser parameters are $I=6\times10^{14}$ W$\cdot$cm$^{-2}$ and $\lambda=800$ nm. We have used a trapezoidal shaped pulse with 3 optical cycles 'turn-on' and 'turn-off' and a plateau of 4 constant-amplitude optical cycles (see panel (d) of Fig. 9). The arrow indicates the cutoff predicted by the semiclassical model~\cite{sfa}. Panel (a) homogeneous case, (b) $\varepsilon=0.01$ (100 a.u), (c) $\varepsilon=0.02$ (50 a.u) and (d) $\varepsilon=0.05$ (20 a.u). The zoomed regions correspond to harmonic order values between $100\;\omega$ and $120\;\omega$ in panel (a) and $330\;\omega$ and $350\;\omega$ in panel (d), respectively (see text for further details).}
%\label{fig:figure8}
%\end{figure}
%
%
%\begin{figure}[htb]
%\centering
%\includegraphics[width=\textwidth]{figure9.eps}
%\caption{Panels (a)-(d): Gabor analysis for the HHG spectra of Figure 8. The zoomed regions in all panels show a time interval during the laser pulse for which the complete electron trajectory, from birth time to recollision time, falls within the pulse plateau (see text and Ref.~\cite{manfred} for details); panel (e) shape of the laser electric field.  In panels (a)-(d) the color scale is logarithmic.}
%\label{fig:figure9}
%\end{figure}

We now concentrate our efforts in to explain, using the semiclassical three-step model, the extension of the harmonic cutoff. As was already pointed out, this new feature appears as a consequence of the combination two factors, namely the nonhomogeneous character of the laser electric field and the electron confinement. It is well known that the position of the HHG cutoff holds 
\begin{equation}
\label{cutoff}
n_{c}\omega=3.17 U_p+I_p
\end{equation}
where $n_c$ is the harmonic order at the cutoff, $\omega$ the laser frequency, $U_p$ the ponderomotive energy ($U_p=I/4\omega^2$, $I$ being the laser intensity in a.u.) and $I_p$ the ionization potential of the atom or molecule~\cite{sfa}. This relationship can be obtained solving the classical Newton equation for an electron moving in a linearly polarized electric oscillating field under the following conditions (the resulting model is also known as the simple man's model): (i) the electron starts with zero velocity at position zero at time $t = t_0$ ($t_0$ is known as ionization time), i.e. $x(t_0)=0$ and $\dot{x}(t_0)=0$ for our 1D model; (ii) when the electric field reverses its direction, the electron returns to its initial position (i.e \textit{recollides} with its parent ion) at a later time $t=t_1$ ($t_1$ is also known as recollision time), i.e. $x(t_1)=0$. The electron kinetic energy at the return time $t_1$ can be obtained from $E_k(t_1)=\dot{x}(t_1)^{2}/2$ and finding the value of $t_1$ (as a function of $t_0$) that maximizes this energy, Eq. (\ref{cutoff}) is fulfilled.

%\begin{figure}[htb]
%\centering
%\includegraphics[width=0.6\textwidth]{figure10.eps}
%\caption{Total energy of the free electron (in terms of the harmonic order) in the laser field when it recollides with its parent ion obtained from Newton's second law and plotted as a function of the ionization time (green open circles) or the recollision time (red filled circles). Panel (a) homogeneous case, (b) $\varepsilon=0.01$ (100 a.u), (c) $\varepsilon=0.02$ (50 a.u) and (d) $\varepsilon=0.05$ (20 a.u). In all the cases the motion of the electron is not restricted.}
%\label{fig:figure10}
%\end{figure}

We have solved numerically the Newton equation for an electron moving in a linearly polarized (in the $x$-axis) electric field with the same parameters used in the 1D-TDSE calculations. Specifically we find the numerical solution of $\ddot{x}(t)=-\nabla_{x}V_{laser}(x,t)$ with $V_{laser}(x,t)$ given by (\ref{vlaser}) and $E(x,t)$ of Eqs. (\ref{electric}) and (\ref{shape}) without any approximation, i.e. we solve the Eq. (\ref{newton}).  For fixed values of ionization time $t_0$ it is possible to compute the classical trajectories and to numerically calculate the times $t_1$ where the electron recollides, i.e. $x(t_1)=0$. Also, once the ionization time $t_0$ is fixed, the electron trajectory is completely determined. In Fig. 10 panels (a)-(d) we plot the dependence of the harmonic order on the ionization time ($t_0$) and recollision time ($t_1$), calculated from $n=(E_k(t_i)+I_p)/\omega$, with $i=0$ and $i=1$, and for the cases presented in the 1D-TDSE simulations, i.e. homogeneous (panel a),  $\varepsilon=0.01$ (panel b), $\varepsilon=0.02$ (panel c) and $\varepsilon=0.05$ (panel d), respectively. At this point we have not restricted the electron trajectories, and consequently we allow the electron to move in all the space (see below for details). The temporal axis, i.e. the x-axis, is plotted in terms of optical cycles and we have chosen a \textit{temporal} window from 3.5 to 5 optical cycles (i.e. from 380 a.u. to 550 a.u.). In all the panels both the short and long trajectories (see see e.g.~\cite{yavuz}) are labeled. From panel (a) it is possible to observe that the maximum kinetic energy of the returning electron is in perfect agreement with Eq. (\ref{cutoff}) (no harmonic order beyond $n_c\sim 36$ is reached). On the other hand panels (b)-(d) show how the nonhomogeneous field modifies the electron trajectories and that no clear HHG cutoff is observed. This is in consistent with the predictions of the 1D-TDSE for the largest spatial grids (see Figs. (2) and (4)). Although from panels (b)-(d) we observe a no clear HHG cutoff, the kinetic energy gained by the electron is finite in agreement with the energy conservation. Furthermore, panels (b)-(d) show similar features as those observed in Fig. (2), panels (b)-(d) and Fig. (4), panels (b)-(d), i.e. only \textit{extended} short trajectories contribute to the harmonic radiation. This characteristic is related with the modifications the electron trajectories suffer due to the nonhomogeneities of the field (see below for details).

%\begin{figure}[htb]
%\centering
%\includegraphics[width=0.6\textwidth]{figure11.eps}
%\caption{Idem Fig. 10, but with the motion of the electron confined into a region $[-\alpha_0,+\alpha_0]$.}
%\label{fig:figure11}
%\end{figure}

In order to classically simulate the 1D-TDSE results, but for the smallest grid size, i.e. $x_{lim}=\pm 1.5 \alpha_0$, we restrict the classical electron trajectories to the domain $[-\alpha_0,\alpha_0]$. The $\pm \alpha_0$ values represent the starting point of the mask function and consequently a fair comparison is possible. The results are presented in Fig. 11, panels (a)-(d). From these plots we can argue that only short trajectories contributes to the harmonic radiation. This is related with the electron motion restriction, i.e. the \textit{confinement}, we have incorporated in the classical simulations. Furthermore, a clear HHG cutoff is now observed for all the nonhomogenous cases and its value is in clear agreement with the 1D-TDSE predictions (see Figs. (5) and (6)).

%\begin{figure}[htb]
%\centering
%\includegraphics[width=0.6\textwidth]{figure12.eps}
%\caption{Dependence of the semiclassical trajectories on the ionization and recollision times for different values of $\epsilon$ and for the non confined case, panel (a) and the confined case, panel (b). Red squares ($\color{red}\Box$), homogeneous case $\varepsilon=0$; green circles ($\color{green}\bigcirc$) $\varepsilon=0.01$; blue triangles ($\color{blue}\triangle$) $\varepsilon=0.02$ and blue triangles ($\color{magenta}\diamond$) $\varepsilon=0.05$.}
%\label{fig:figure12}
%\end{figure}

Finally,  in Fig. 12 the recollision time $t_1$ of the electron is presented as a function of the ionization time $t_0$ and for several values of $\varepsilon$. Panel (a) represents the non-confined case and in panel (b) we have restricted the electron motion into the region $[-\alpha_0,\alpha_0]$. The long trajectories are those with recollision times $t_1\gtrsim4.25$ optical cycles and only for the homogeneous case (red squares ($\color{red}\Box$)) these trajectories are clearly visible. On the other hand, short trajectories are characterized by $t_1\lesssim4.25$ optical cycles and these are present for both the homogeneous and nonhomogeneous cases. Our results are consistent with those shown in~\cite{yavuz}, but note our inhomogeneity parameter is more than one order of magnitude higher. From panel (a) we observe how the long trajectories are modified by the nonhomogeneity, namely the \textit{homogeneous} long trajectories (red squares ($\color{red}\Box$)) with ionization times $t_0$ around the 3.25  and 3.75 optical cycles \textit{merge} into unique trajectories. Additionally the branch with $t_0\sim 3.75$ has now ionization times more than half an optical cycle smaller when $\varepsilon$ increases; hence, the times spent by the electron in the continuum increase. 
This fact explains the vanishing long trajectories seen in the panels (b)-(d) of Fig. 10. Additionally, the electric field strength at the ionization time for short trajectories is higher than for long trajectories and considering the ionization rate as a highly nonlinear function of this electric field~\cite{keldysh65,ammosov87}, long trajectories are much less efficient than the short ones. On the other hand, short trajectories are almost independent of $\varepsilon$ and only for higher values noticeable differences are visible. When the electron motion is confined, panel (b), only short trajectories are present for all the cases and this confirm the fact that long trajectories are absent and only the short trajectories contribute to the formation of the harmonic radiation.

\section{Conclusions and Outlook}

We studied high-order harmonic generation in a model atom produced by non-homogeneous fields. These fields are not merely a theoretical speculation but are present in a vicinity of a metal nanostructure when it is irradiated by a short laser pulse. We have extended two of the most widely used  models, namely the numerical solution of the time dependent Schr\"odinger equation (TDSE) in reduced dimensions and the semiclassical approach known as Strong Field Approximation (SFA). Both models are able to predict the new features that appear due to presence of inhomogeneities, namely the appearance of odd and even harmonics and an extension in the cutoff position. The latter feature is a consequence of the combination of the nonhomogeneous electric field and the confinement of the electron dynamics as can be extracted from the Gabor analysis and the semiclassical simulations made for all the studied cases. We have proved the robustness of the 1D-TDSE model using different laser parameters and solving the classical equation of motion for an electron moving in a linearly polarized nonhomogeneous electric field we are able to understand the reasons of the cutoff extension.

The models presented allow to use, in principle, any functional form for the nonhomogeneous fields. Moreover they are able to incorporate real parameters in the simulations, such as the region dimensions where the electron dynamics takes place. We plan to continue our investigations in this direction, joint with studies of other laser-matter processes driven now by these  nonhomogeneous fields.

\section*{Acknowledgments}

We acknowledge the financial support of the MINCIN projects (FIS2008-00784 TOQATA and Consolider Ingenio 2010 QOIT) (M. F. C. and M.L.); ERC Advanced Grant QUAGATUA, Alexander von Humboldt Foundation and Hamburg Theory Prize (M. L.); Spanish Ministry of Education and Science through its Consolider Program Science (SAUUL CSD 2007-00013), Plan Nacional (FIS2008-06368-C02-01), LASERLAB-EUROPE (grant agreement n° 228334, EC's Seventh Framework Programme) (J. B.); this research has been partially supported by Fundaci\'o Privada Cellex. Useful and enlightening discussions with Manfred Lein are gratefully acknowledged. M.F.C. acknowledges Mitsuko Korobkin and Dane Austin for help and advices in the numerical implementation of the 1D-TDSE model.

%\bibliography{plasmonics}

\newpage
Figures captions \newline

Fig. 1. High-order harmonic generation (HHG) spectra for a model atom with $\mathcal{E}_{GS}=-0.67$ a.u. generated using the 1D-TDSE model and with a spatial grid of $x_{lim}=\pm7.5 \alpha_0$ (see text for details). The laser parameters are $I=2\times10^{14}$ W$\cdot$cm$^{-2}$ and $\lambda=800$ nm. We have used a trapezoidal shaped pulse with two optical cycles turn on, i.e. $n_{on}=2$, and turn off, i.e. $n_{off}=2$, and a plateau with six optical cycles, i.e. $n_{p}=6$ (10 optical cycles in total, i.e. approximately 27 fs). The arrow indicates the cutoff predicted by the semiclassical model~\cite{sfa}. Panel (a) homogeneous case, (b) $\varepsilon=0.01$ (100 a.u), (c) $\varepsilon=0.02$ (50 a.u) and (d) $\varepsilon=0.05$ (20 a.u).

Fig. 2. (Color online) Panels (a)-(d): Gabor analysis for the HHG spectra of Figure 1. The zoomed regions in all panels show a time interval during the laser pulse for which the complete electron trajectory, from birth time to recollision time, falls within the pulse plateau (see Ref.~\cite{manfred} for details); panel (e) shape of the laser electric field. In panels (a)-(d) the color scale is logarithmic.

Fig. 3. Idem Fig. 1 but with a spatial grid of $x_{lim}=\pm4.5 \alpha_0$.

Fig. 4. (Color online) Gabor analysis for the HHG spectra of Figure 3. The zoomed regions in all panels show a time interval during the laser pulse for which the complete electron trajectory, from birth time o recollision time, falls within the pulse plateau (see text and Ref.~\cite{manfred} for details).

Fig. 5. Idem Fig. 1 but with a spatial grid of $x_{lim}=\pm1.5 \alpha_0$.

Fig. 6. (Color online) Gabor analysis for the HHG spectra of Figure 5. The zoomed regions in all panels show a time interval during the laser pulse for which the complete electron trajectory, from birth time o recollision time, falls within the pulse plateau (see text and Ref.~\cite{manfred} for details).

Fig. 7. HHG spectra for a model atom with $\mathcal{E}_{GS}=-0.67$ a.u. generated using the 1D-TDSE model  and with a spatial grid of $x_{lim}=\pm1.5 \alpha_0$ (see text for details). The laser parameters are $I=1\times10^{13}$ W$\cdot$cm$^{-2}$ and $\lambda=3.2$ $\mu$m (3200 nm). We have used a gaussian shaped pulse with 6 cycles FWHM. The arrow indicates the cutoff predicted by the semiclassical model~\cite{sfa}. The insets in panels (a) and (d) show particular zoomed regions of the harmonic spectra near the cutoff region (see the text for details).

Fig. 8. (Color online) HHG spectra for a model atom with $\mathcal{E}_{GS}=-0.67$ generated using the extended SFA approach. The laser parameters are $I=6\times10^{14}$ W$\cdot$cm$^{-2}$ and $\lambda=800$ nm. We have used a trapezoidal shaped pulse with 3 optical cycles 'turn-on' and 'turn-off' and a plateau of 4 constant-amplitude optical cycles (see panel (d) of Fig. 9). The arrow indicates the cutoff predicted by the semiclassical model~\cite{sfa}. Panel (a) homogeneous case, (b) $2\varepsilon=0.01$ (100 a.u), (c) $2\varepsilon=0.02$ (50 a.u) and (d) $2\varepsilon=0.05$ (20 a.u). The zoomed regions correspond to harmonic order values between $100\;\omega$ and $120\;\omega$ in panel (a) and $330\;\omega$ and $350\;\omega$ in panel (d), respectively (see text for further details).

Fig. 9.(Color online)  Panels (a)-(d): Gabor analysis for the HHG spectra of Figure 8. The zoomed regions in all panels show a time interval during the laser pulse for which the complete electron trajectory, from birth time to recollision time, falls within the pulse plateau (see text and Ref.~\cite{manfred} for details); panel (e) shape of the laser electric field.  In panels (a)-(d) the color scale is logarithmic.

Fig. 10. (Color online) Total energy of the free electron (in terms of the harmonic order) in the laser field when it recollides with its parent ion obtained from Newton's second law and plotted as a function of the ionization time (green open circles) or the recollision time (red filled circles). Panel (a) homogeneous case, (b) $\varepsilon=0.01$ (100 a.u), (c) $\varepsilon=0.02$ (50 a.u) and (d) $\varepsilon=0.05$ (20 a.u). In all the cases the motion of the electron is not restricted.

Fig. 11. (Color online) Idem Fig. 10, but with the motion of the electron confined into a region $[-\alpha_0,+\alpha_0]$.

Fig. 12. (Color online) Dependence of the semiclassical trajectories on the ionization and recollision times for different values of $\epsilon$ and for the non confined case, panel (a) and the confined case, panel (b). Red squares ($\color{red}\Box$), homogeneous case $\varepsilon=0$; green circles ($\color{green}\bigcirc$) $\varepsilon=0.01$; blue triangles ($\color{blue}\triangle$) $\varepsilon=0.02$ and magenta diamonds ($\color{magenta}\diamond$) $\varepsilon=0.05$.

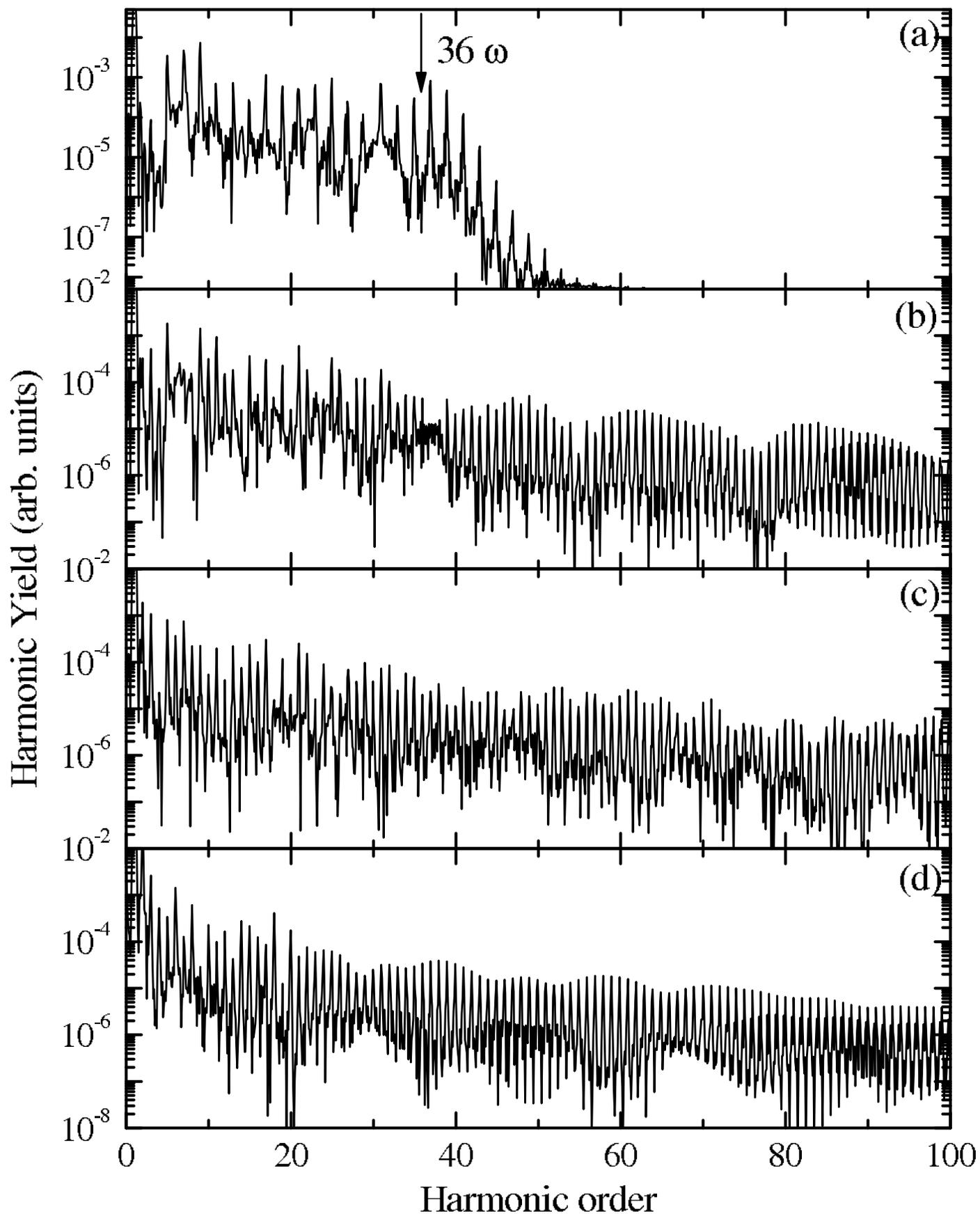
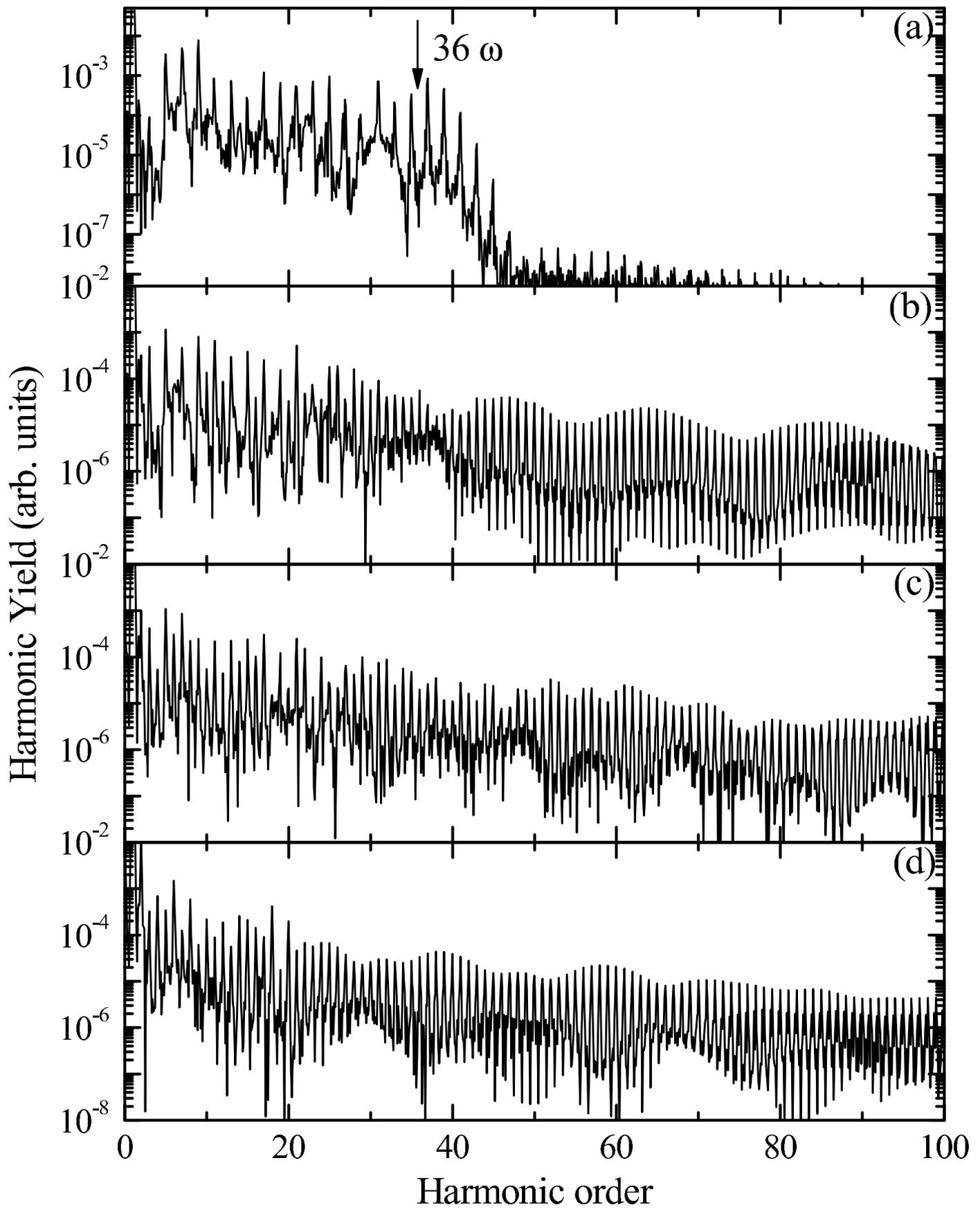
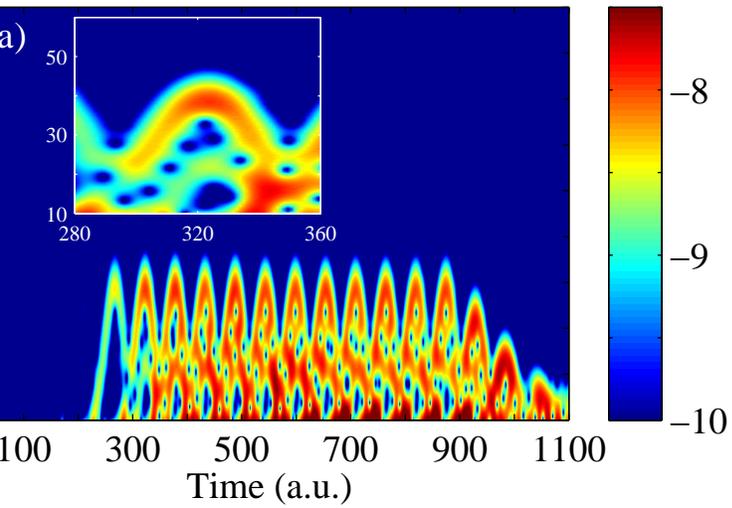
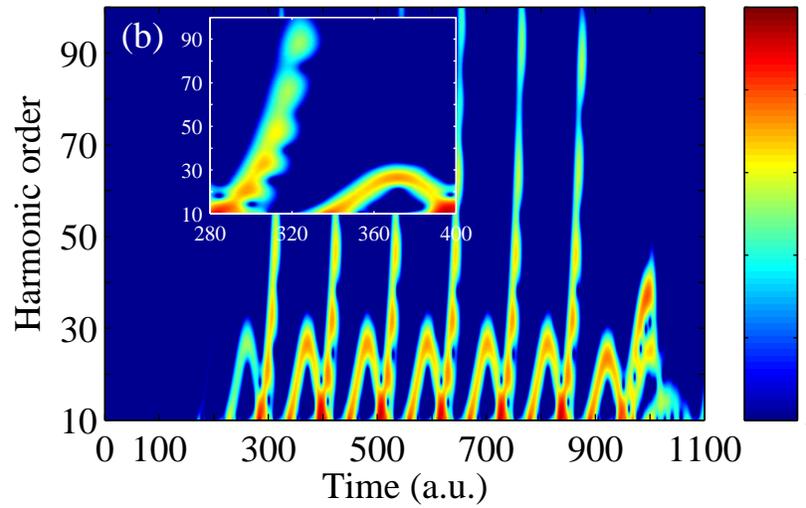
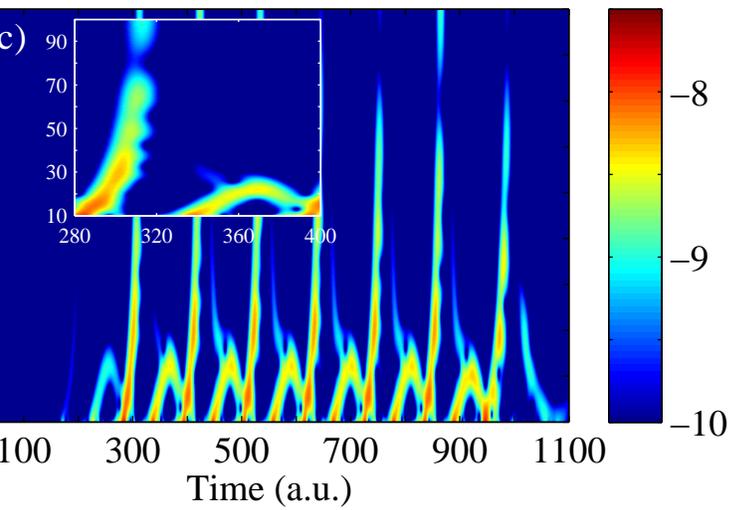
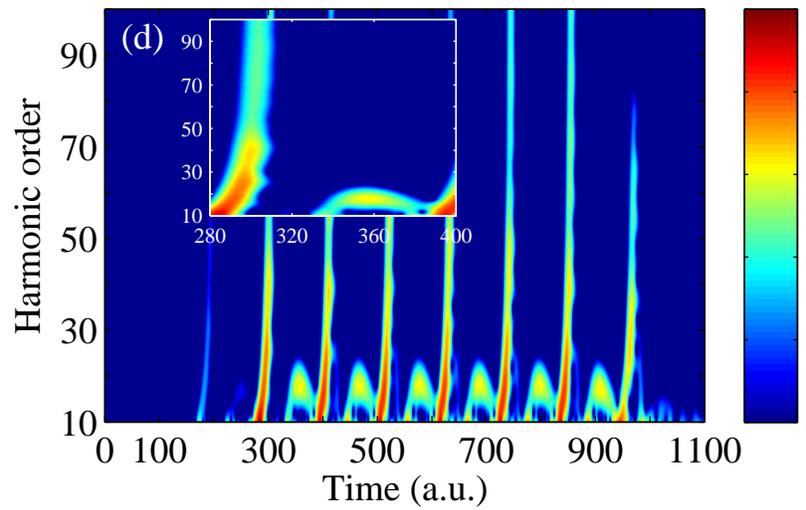
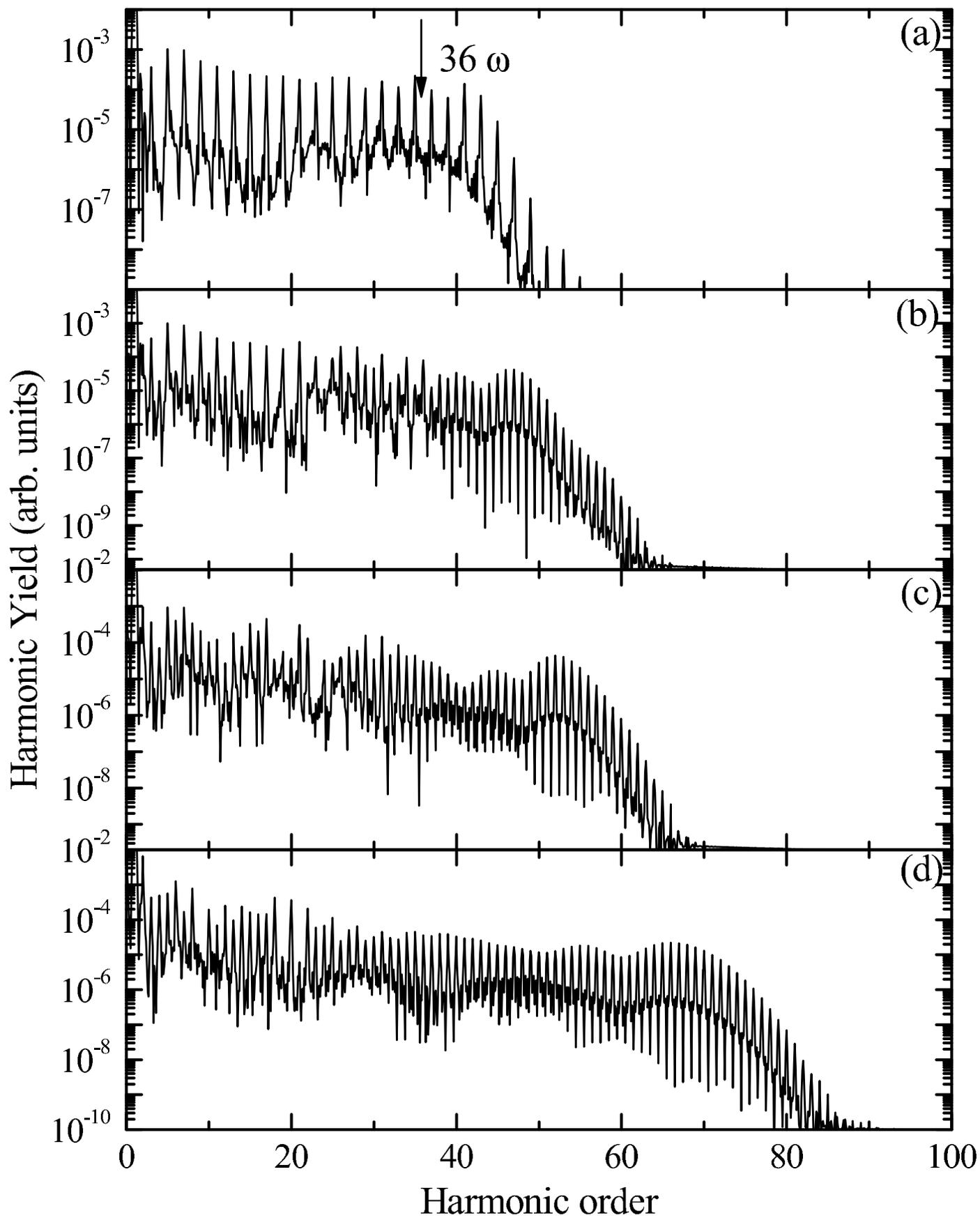
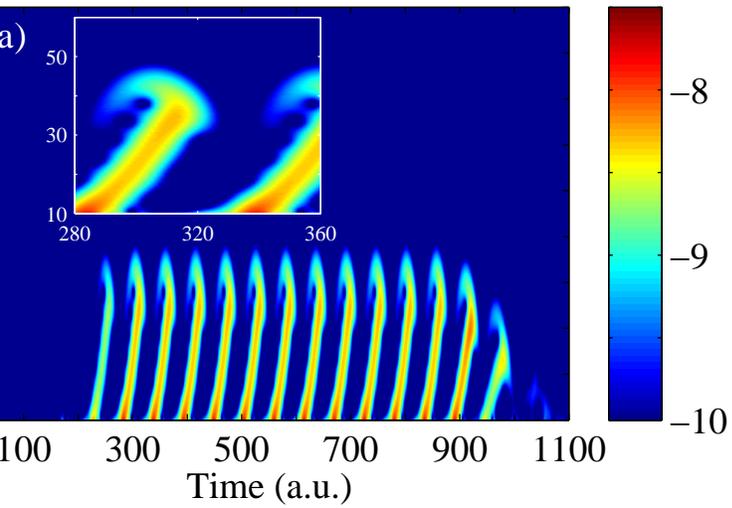
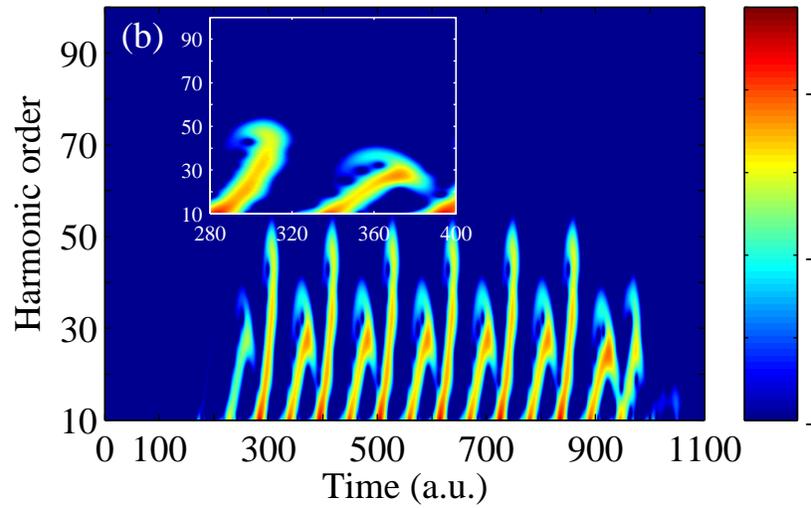
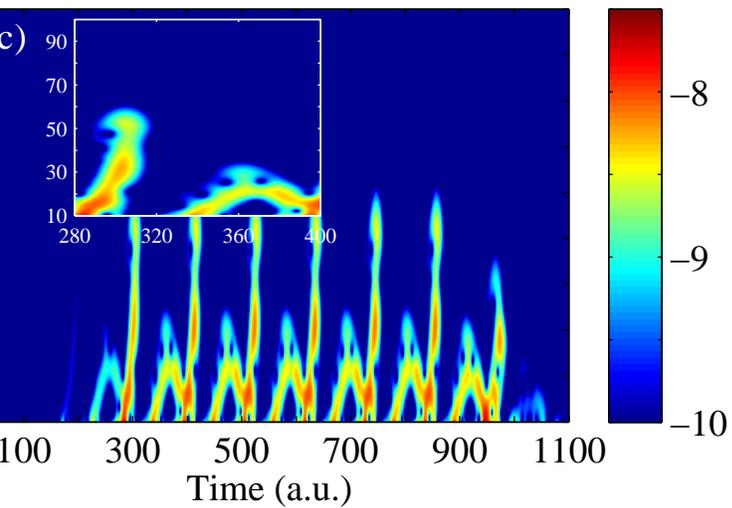
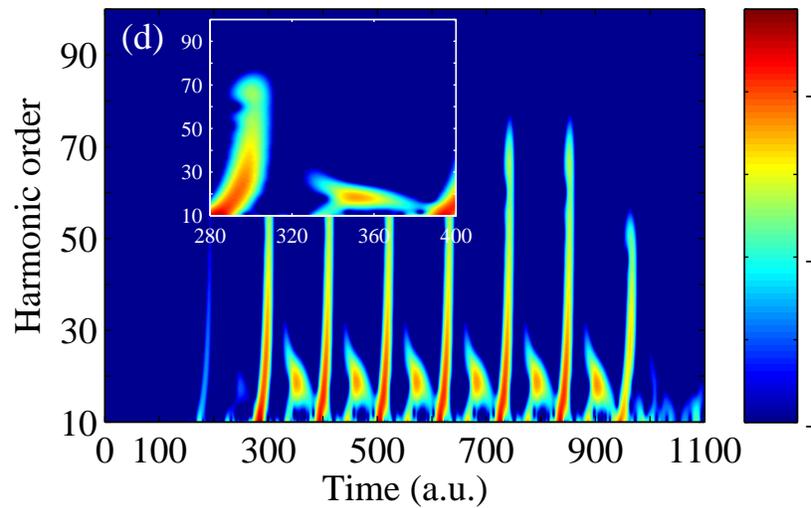
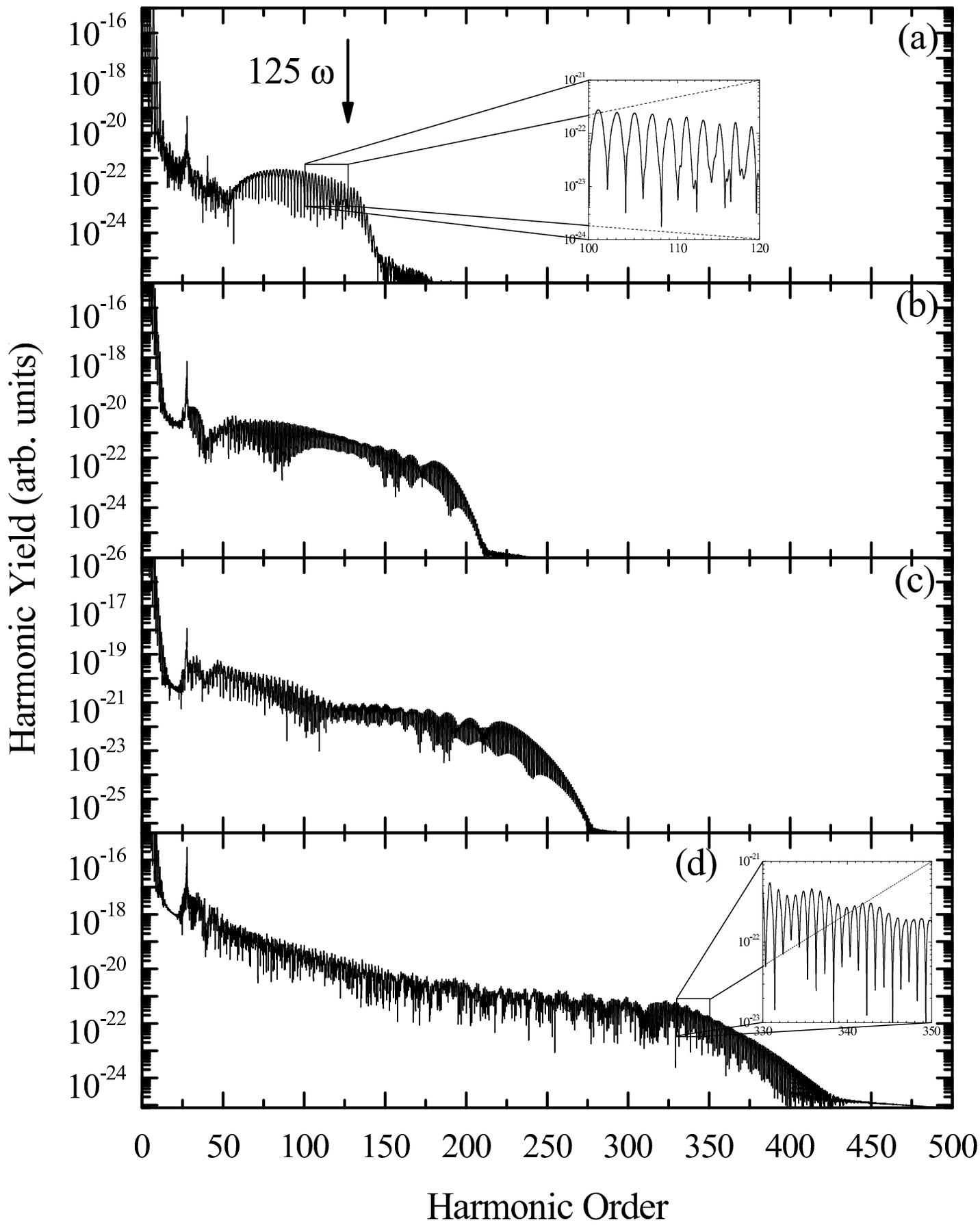
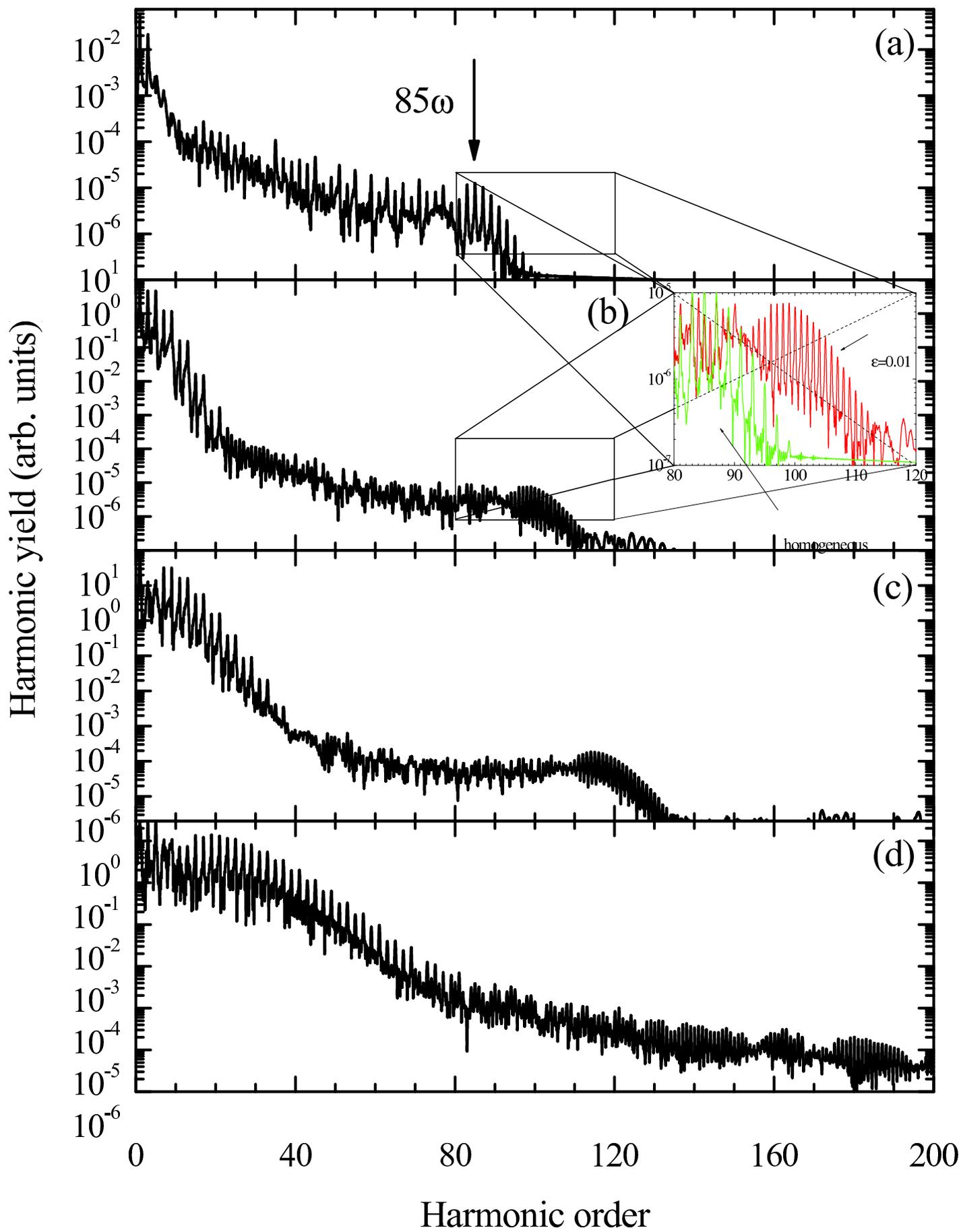
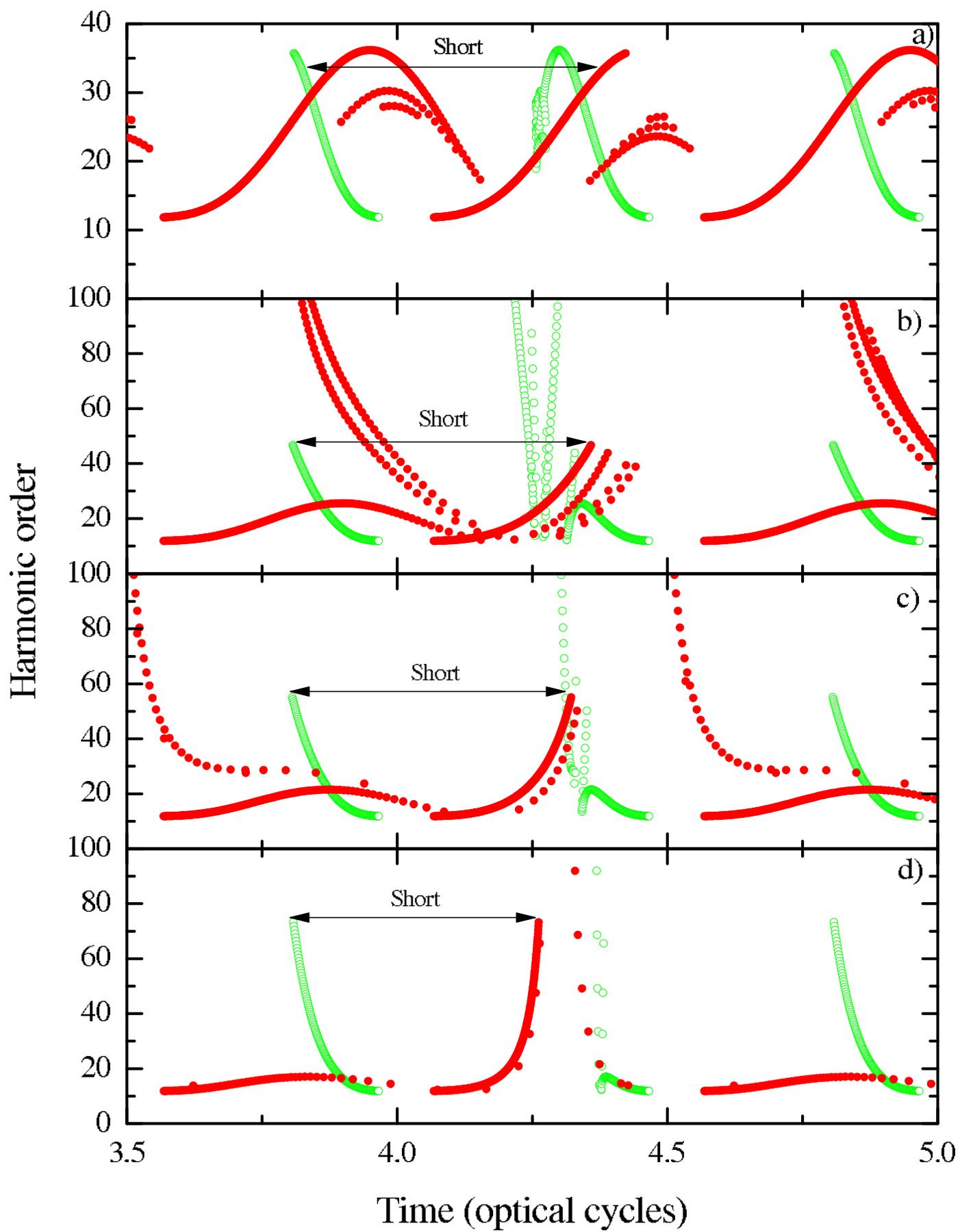
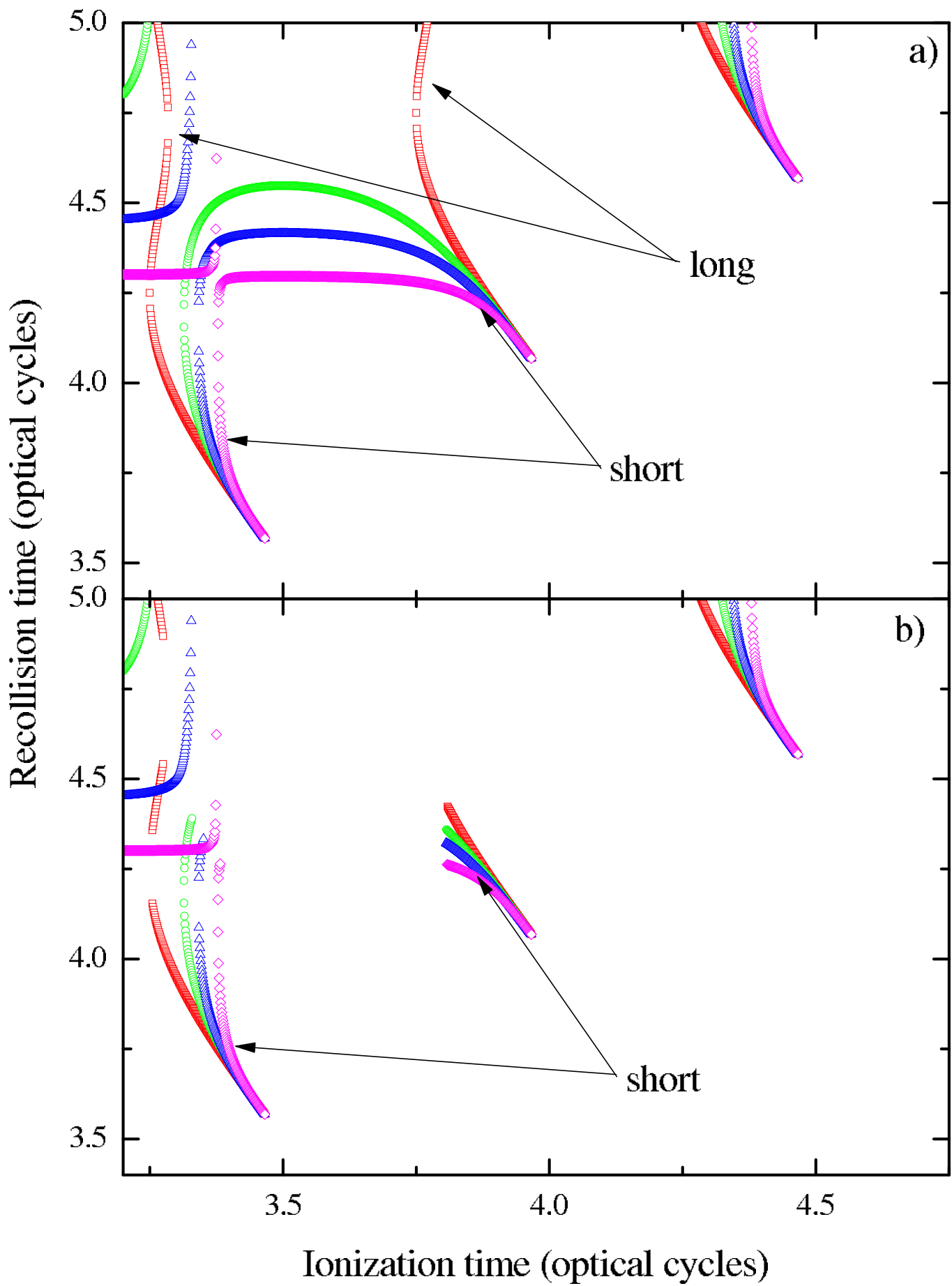
\end{document}